\documentclass[aps,pre,twocolumn,groupedaddress,superscriptaddress,showpacs]{revtex4-1}
\usepackage{amsmath}
\usepackage{amssymb}
\usepackage{graphicx}
\usepackage{float}
\usepackage{color}

\begin{document}

\title{Dynamics of patchy particles in and out of equilibrium}

  \author{J. M. Tavares}
   \email{jmtavares@fc.ul.pt}
    \affiliation{Centro de F\'{i}sica Te\'{o}rica e Computacional, Universidade de Lisboa, 
    1749-016 Lisboa, Portugal}
    \affiliation{Instituto Superior de Engenharia de Lisboa, ISEL, Avenida Conselheiro Em\'{i}dio Navarro, 1  1950-062 Lisboa, Portugal}
    
  \author{C. S. Dias}
   \email{csdias@fc.ul.pt}
    \affiliation{Centro de F\'{i}sica Te\'{o}rica e Computacional, Universidade de Lisboa, 
    1749-016 Lisboa, Portugal}
    \affiliation{Departamento de F\'{\i}sica, Faculdade de Ci\^{e}ncias, Universidade de Lisboa, 
    1749-016 Lisboa, Portugal}

  \author{N. A. M. Ara\'ujo}
   \email{nmaraujo@fc.ul.pt}
    \affiliation{Centro de F\'{i}sica Te\'{o}rica e Computacional, Universidade de Lisboa, 
    1749-016 Lisboa, Portugal}
    \affiliation{Departamento de F\'{\i}sica, Faculdade de Ci\^{e}ncias, Universidade de Lisboa, 
    1749-016 Lisboa, Portugal}
    
  \author{M. M. Telo da Gama}
   \email{mmgama@fc.ul.pt}
    \affiliation{Centro de F\'{i}sica Te\'{o}rica e Computacional, Universidade de Lisboa, 
    1749-016 Lisboa, Portugal}
    \affiliation{Departamento de F\'{\i}sica, Faculdade de Ci\^{e}ncias, Universidade de Lisboa, 
    1749-016 Lisboa, Portugal}

\begin{abstract}
We combine particle-based simulations, mean-field rate equations, and
Wertheim's theory to study the dynamics of patchy particles in and out of
equilibrium, at different temperatures and densities. We consider an initial
random distribution of non-overlapping three-patch particles, with no bonds,
and analyze the time evolution of the breaking and bonding rates of a single
bond. We find that the asymptotic (equilibrium) dynamics differs from the
initial (out of equilibrium) one. These differences are expected to depend on
the initial conditions, temperature, and density.
\end{abstract}

\maketitle

\section{Introduction}
The possibility of synthesizing novel materials with enhanced physical
properties from the spontaneous self-assembly of colloidal particles is among
the most popular challenges of Soft Condensed
Matter~\cite{Doppelbauer2010,Glotzer2007,Frenkel2011,Kufer2008,Whitesides2002,Ma2011}.
Functionalized colloidal particles with patches on their surfaces (patchy
particles) are promising candidates for the individual constituents, as they
allow control of both the valence (number of neighboring particles) and the
local
structure~\cite{Zhang2014a,Paulson2015,Lu2013,Duguet2011,Hu2012,Kretzschmar2011,Sacanna2011,Solomon2011,Pawar2010,Sacanna2013a,Manoharan2015}.
Equilibrium studies of patchy particle systems revealed rich phase diagrams depending
on temperature and density, and in novel ways on the number and type of
patches~\cite{Zaccarelli2007,Kretzschmar2011,Dias2017}. However, the potential
application of these predictions may be compromised, as the feasibility of
assembling the new phases is seriously hampered by the kinetic barriers that
emerge from the complex particle-particle
correlations~\cite{Zaccarelli2006,Dias2013,Dias2013a,Chakrabarti2014,Dias2014,Dias2015,Araujo2015,Kartha2016,Dias2016,Kartha2016a,Araujo2017}.
Thus, understanding the dynamics towards thermodynamically stable (equilibrium)
structures is vital to tackle this challenge.

Previous studies attempted to relate the out of equilibrium dynamics to
the dynamics at equilibrium~\cite{Corezzi2010,Corezzi2009,Sciortino2009}. The
idea is to describe the dynamics as a balance between breaking and forming
individual bonds, where the rates for each process are considered time
invariant. Since the overall bond probability should converge to that at
equilibrium, it is possible to estimate the effective rates from a fit to the
time evolution of the bond probability. These studies were restricted to
systems where the structures are almost loopless and characterized by very long
chains. In this work, we consider systems of three-patch particles, where
branching and loops are significant. We combine particle-based (Langevin)
simulations, mean-field rate equations, and Wertheim's theory to study the
dynamics in and out of equilibrium. We show that, while the effective rates
obtained from the time evolution of the bond probability describe the out of 
equilibrium dynamics accurately, the equilibrium dynamics is significantly different and
characterized by higher (time invariant) rates.

The manuscript is organized in the following way. In the next section we
present the model of patchy particles and describe the simulations. In
Sec.~\ref{sec.results}, we introduce the mean-field approach and compare the
equilibrium and out of equilibrium regimes combining Wertheim's theory and
particle-based simulations. Finally, we draw some conclusions in
Sec.~\ref{sec.conclusions}.

\section{Model and simulations\label{sec.model}}
Following previous works~\cite{Dias2016,Dias2016a,Dias2017b}, we model the
patchy particles as three-dimensional spheres, with three patches equally
distributed on their surfaces. The core-core interaction is described by a
(repulsive) Yukawa-like potential,
$V_Y(r)=\frac{A}{k}\exp{\left(-k\left[r-\sigma\right]\right)}$, where $\sigma$ 
is the effective diameter of the interacting particles, $A/k=0.25k_BT$ is the
interaction strength and $k/\sigma=0.4$ is the inverse screening length. The patch-patch
interaction is described by an attractive pairwise potential
$V_G(r_p)=-\epsilon\exp\left[-(r_p/\xi)^2\right]$, where $\epsilon$ is the
interaction strength, $\xi=0.1\sigma$ is the size of the patch, and $r_p$ the distance
between patches~\cite{Vasilyev2013}.

To resolve the individual trajectories of the particles, we perform
particle-based simulations using the velocity Verlet scheme of the Large-scale
Atomic/Molecular Massively Parallel Simulator (LAMMPS)~\cite{Plimpton1995}. We
integrate the Langevin equations of motion for the translational and
rotational degrees of freedom, respectively,
\begin{equation}
 m\dot{\vec{v}}(t)=-\nabla_{\vec{r}}
 V(\vec{r})-\frac{m}{\tau_t}\vec{v}(t)+\sqrt{\frac{2mk_BT}{\tau_t}}\vec{f_t}(t),
 \label{eq.Langevin_dynamics_trans} 
\end{equation}
and
\begin{equation}
	I\dot{\vec{\omega}}(t)=-\nabla_{\vec{\theta}}
	V(\vec{\theta})-\frac{I}{\tau_r}\vec{\omega}(t)+\sqrt{\frac{2Ik_BT}{\tau_r}}\vec{f_r}(t),
	\label{eq.Langevin_dynamics_rot}
\end{equation}
where, $\vec{v}$ and $\vec{\omega}$ are the translational and angular
velocities, $m$ and $I$ are the mass and moment of inertia of each patchy
particle, $V(\cdots)$ is the pairwise potential, $\tau_t$ and $\tau_r$ are the
translational and rotational damping times, and $\vec{f_t}(t)$ and
$\vec{f_r}(t)$ are stochastic terms taken from a random distribution of zero
mean.  The damping time for the translational
motion is $\tau_t=m/(6\pi\eta R)$, and the damping
time for the rotational motion is $\tau_r=10\tau_t/3$, in line with
the Stokes-Einstein-Debye relation for spheres~\cite{Mazza2007}.

Simulations are performed for a three-dimensional system of linear size
$L=16$, in units of the particle diameter $\sigma$, averaged over 20 samples, starting from a random
(uniform) distribution of non-overlapping particles. The density $\rho$ is the 
number of particles per unit volume. We consider five different
densities, namely, $\rho=\{\frac{1}{64}, \frac{1}{32},
\frac{1}{16}, \frac{1}{8}, \frac{1}{4}\}$ in units of $1/\sigma^3$, and four
different temperatures: $T=\{0.075, 0.1, 0.125, 0.15\}$ in units of $\frac{\epsilon}{k_B}$.

\section{Results\label{sec.results}}
%

%%%%%%%%%%%%%%%%
\begin{figure}
   \begin{center} \includegraphics[width=\columnwidth]{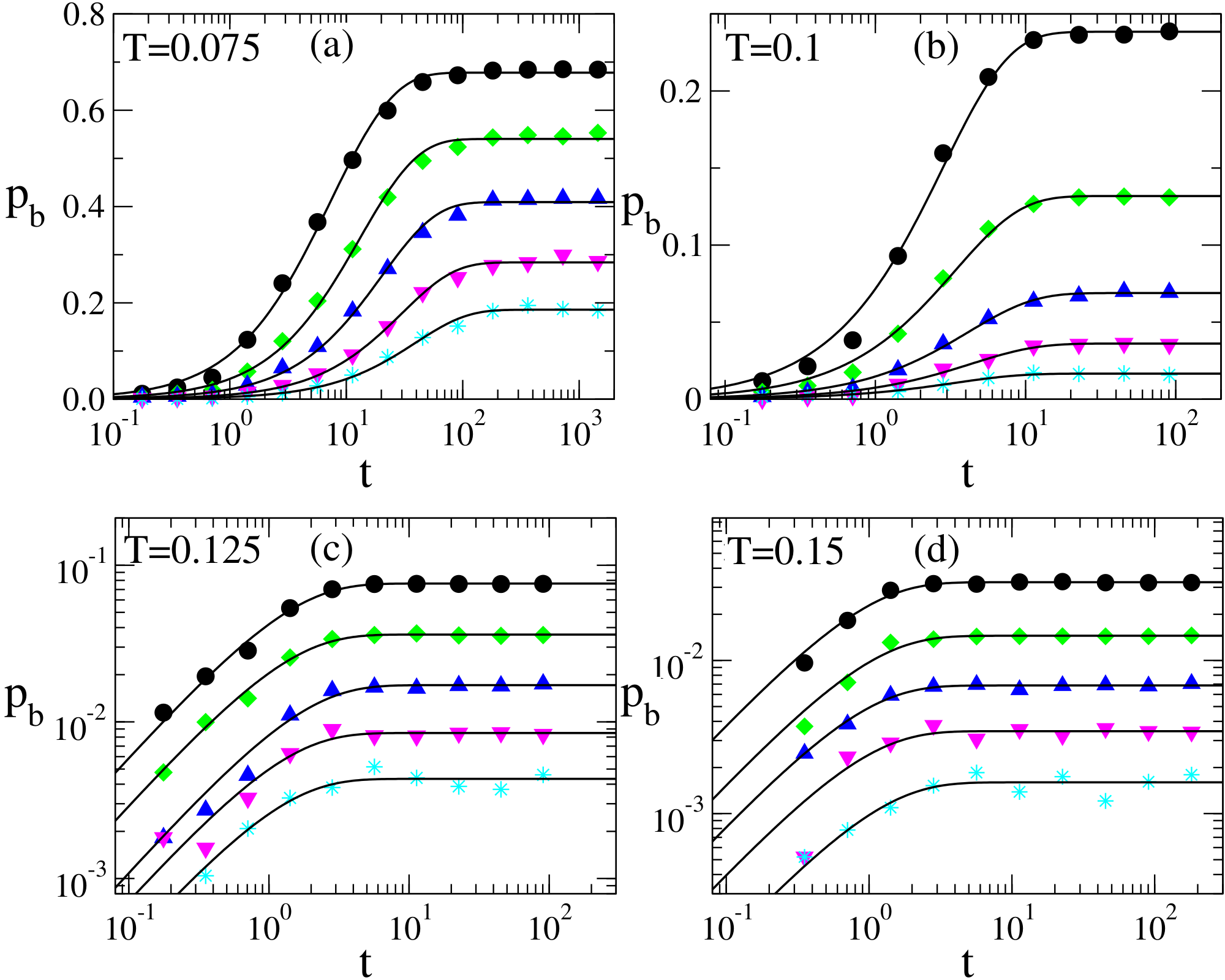} \\
	   \caption{Time dependence of the bonding probability, $p_b(t)$,
		   obtained from the particle-based simulations (symbols) at
		   different densities, namely, $\frac{1}{64}$ (stars),
		   $\frac{1}{32}$ (triangles down), $\frac{1}{16}$ (triangles
		   up), $\frac{1}{8}$ (diamonds), $\frac{1}{4}$ (circles).
	   Different plots are for different temperatures: (a) $0.075$; (b)
   $0.1$; (c) $0.125$; (d) $0.15$. The solid lines are fits to Eq.~(\ref{pbt})
   (details in the text).~\label{pbvstime}}
   \end{center}
\end{figure}
%%%%%%%%%%%%%%%%

The dynamics evolves through a sequence of bonding and breaking events. The
simplest approach to describe this dynamics is through a (mean-field) rate
equation for the bonding probability $p_b(t)$, which corresponds to the
fraction of bonded patches in the infinite size limit. Accordingly,
\begin{equation}\label{dpbdt}
	\dot{p_b}(t)=-k_{br} p_b(t)+\rho f k_{bo}[1-p_b(t)]^2,
\end{equation}
where $\dot{p_b}(t)$ stands for the time derivative, $t$ for time in units of
the Brownian time, $f$ for the valence, $\rho$ for the number density, and
$k_{br}$, $k_{bo}$ for the breaking and bonding rates of a single bond,
respectively.  As discussed in Refs.~\cite{VanDongen1984,Sciortino2009},
Eq.~(\ref{dpbdt}) can be derived from a generalized Smoluchowski set of
equations, accounting for both aggregation and fragmentation of loopless
clusters of particles of functionality $f$ (i.e. $f$ patches), under the
assumptions that all patches are identical, the diffusion coefficient is the
same for all clusters, and all patches are unbonded at $t=0$. If, for
simplicity, we consider that $k_{br}$ and $k_{bo}$ are time invariant and
depend on temperature and density only then,
\begin{equation}\label{pbt}
p_b(t)=p_\infty \frac{1-\exp(-\Gamma t)}{1-p_\infty^2\exp(-\Gamma t)},
\end{equation}
where $p_\infty=\lim_{t\to \infty}p_b(t)$. At equilibrium, $\dot{p_b}(t)=0$,
and the net rates of bonding and breaking are necessarily the same.
Thus,
\begin{equation}\label{gamma}
 \Gamma=k_{br}\frac{1+p_\infty}{1-p_\infty}=f\rho
 k_{bo}\frac{1-p_\infty^2}{p_\infty}.
\end{equation}
Clearly, $p_\infty$ depends only on the ratio $k_{br}/k_{bo}$ and is simply the
bond probability under equilibrium conditions, as discussed below.
The value of $\Gamma$ (and thus the individual values of $k_{br}$ and
$k_{bo}$) may be estimated from a fit of Eq.~(\ref{pbt}) to numerical data
for $p_b(t)$ obtained from particle-based simulations.

%%%%%%%%%%%%%%%%%%%%%%%% 
\begin{figure}
   \begin{center} \includegraphics[width=\columnwidth]{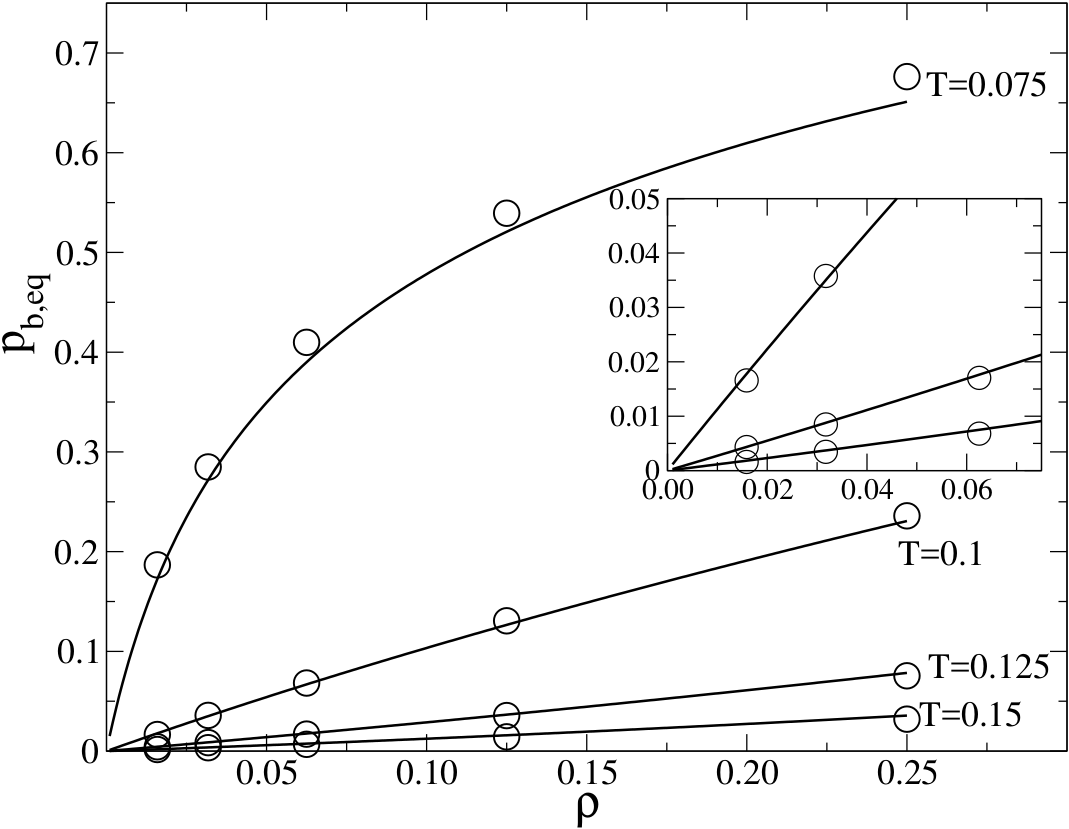} \\
\caption{Density dependence of the equilibrium bonding probability at different
	temperatures. Symbols are estimated from the asymptotic value of
	$p_\infty$ obtained from the simulations, while the lines correspond to
	$p_{b,eq}$ obtained from Wertheim's theory. The behavior at low
densites is shown in the inset.~\label{pbvsrho}}
\end{center}
\end{figure}
%%%%%%%%%%%%%%%%%%%%%%%% 

Wertheim's theory is  the most successful thermodynamic perturbation theory
for associating fluids with short-ranged, anisotropic interactions
(see e.g. \cite{Teixeira2017a}). Within this equilibrium theory, the bonding
free energy, treated as a perturbation over that of a reference fluid, is
found to be a function of the equilibrium bonding probability, $p_{b,eq}$,
which in turn depends on the number density, the functionality $f$, the
interaction potential between the patches $V_G$, and the reference fluid, and
it is the solution of
\begin{equation}\label{lma}
  f\rho\Delta= \frac{p_{b,eq}}{(1-p_{b,eq})^2},
\end{equation}
where
\begin{equation}\label{Delta}
\Delta=\frac{1}{(4\pi)^2}\int d\vec r \int d\hat r_1 \int d\hat r_2
\left(\exp\left[-\beta V_G(r_p)\right]-1\right) g_{ref}(r).
\end{equation}
Here, $g_{ref}(r)$ is the pair correlation function of the reference fluid,
$\vec{r}$ is the vector between the centers of the two particles participating
in the bond, and $\hat r_i$ is the unit vector that defines the position of
the bonded patch on particle $i$ relative to the center of that particle. In
order to proceed we replaced the soft-core repulsion $V_Y(r)$, by the repulsion of hard spheres
with a temperature dependent diameter $d$, obtained through the
Barker-Henderson approximation. Thus, $g_{ref}(r)$ is given by
$g_{HS}(r=d,\rho)$, the contact value of the pair correlation function of a
system of hard spheres of diameter $d$ and number density $\rho$ (see e.g.,
Ref.~\cite{Dias2016a}). $\Delta$ is now rewritten as,
\begin{equation}\label{Delta2}
\Delta=g_{HS}(r=d,\rho)G(T,\alpha,\delta),
\end{equation}
where $\alpha=\xi/\sigma$ and $\delta=(1-d/\sigma)$. $G(T,\alpha,\delta)$ is
the integral of the Mayer function of the patch-patch interaction $V_G$ over
the bond volume (see Ref.~\cite{Dias2016a} for further details).

%%%%%%%%%%%%%%%%%%%%%%%%
\begin{figure}
 \begin{center} 
   \includegraphics[width=\columnwidth]{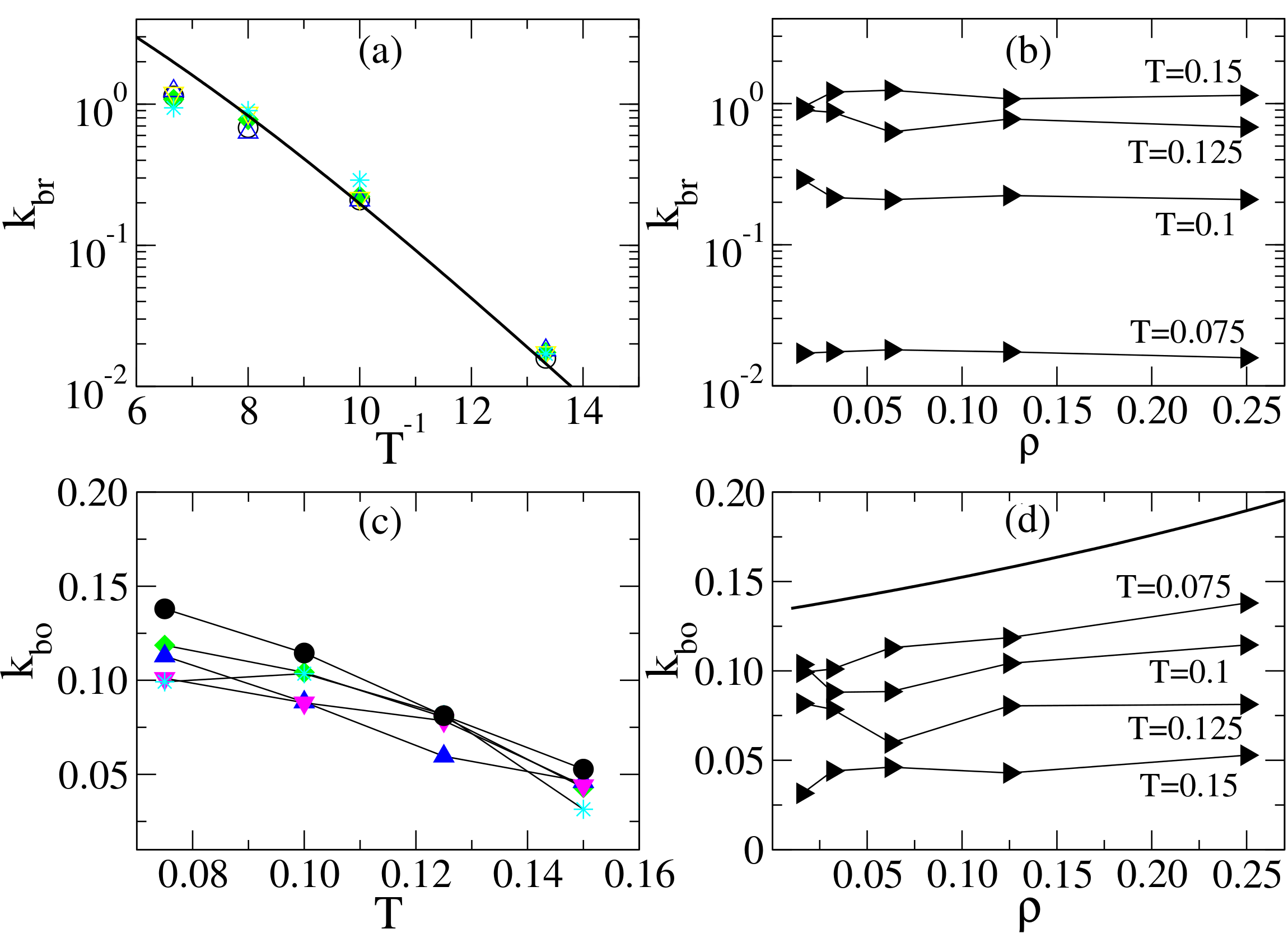} \\
   \caption{Temperature and density dependence of the breaking and bonding
	   rates of a single bond obtained from the fit to the numerical data
	   in Fig.~\ref{pbvstime}. Rate of breaking of a single bond
	   $k_{br}$ as a function of (a) the inverse temperature and (b)
	   density, at different densities and temperatures, respectively. The
	   solid line in (a) is $aG(T)^{-1}$, where $a$ is a constant adjusted
	   to fit the numerical data. Rate of bonding of a single bond $k_{bo}$
   as a function of (c) temperature and (b) density, at different densities and
   temperatures, respectively. The solid line in (d) is proportional to
   $g_{HS}$ at $r=\sigma$ (details in the text).~\label{kbreak}}
\end{center}
\end{figure}
%%%%%%%%%%%%%%%%%%%%%%%%%%%%%%%

Since $p_{b,eq}=p_\infty$, Eqs.~(\ref{dpbdt})~and~(\ref{lma}), imply that at
equilibrium the breaking and bonding rates satisfy,
\begin{equation}\label{Deltakbkb}
 \Delta=\frac{k_{bo}}{k_{br}},
\end{equation} 
consistent with the fact that both Eqs.~(\ref{dpbdt})~and~(\ref{lma}) may be
derived from the Flory-Stockmayer size distribution of
clusters~\cite{VanDongen1984,Bianchi2008}.  It is also known that the
equilibrium mean-field relation breaks down in the limit where a strongly
connected gel is formed~\cite{Corezzi2010}. Note that, both the dynamics and
the equilibrium descriptions, neglect patch-patch correlations. 

%%%%%%%%%%%%%%%%%%%%%%%%%%%%%%%
\begin{figure}
\begin{center} 
  \includegraphics[width=\columnwidth]{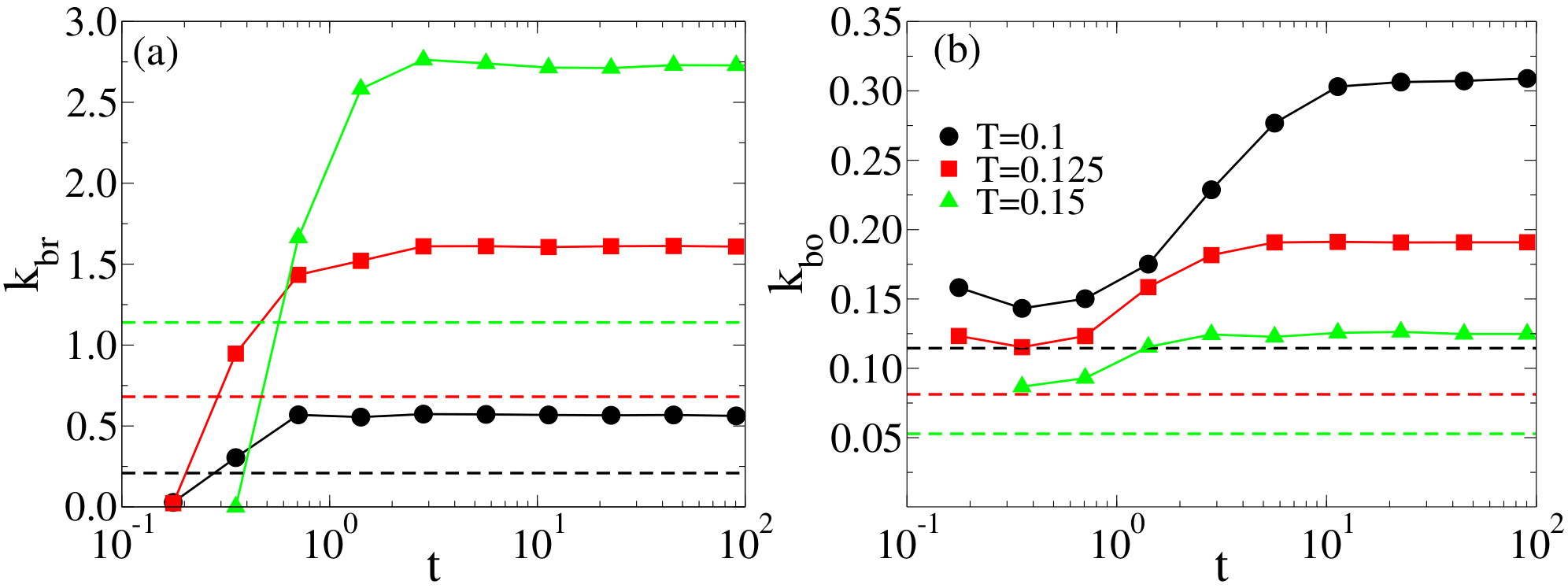} \\
  \caption{Time dependence of the rates of (a) breaking ($k_{br}(t)$) and (b)
  bonding ($k_{bo}(t)$) of individual bonds, obtained from the particle-based
  simulations, at $\rho=0.25$ and three different temperatures: $0.1$; $0.125$;
  $0.15$. The horizontal (dashed) lines correspond to the effective values
  estimated by the fit (details in the text).~\label{kvst}}
\end{center}
\end{figure}
%%%%%%%%%%%%%%%%%%%%%%%%%%%%%%%

In order to assess the validity of the rate equations we compared, in
Fig.~\ref{pbvstime}, the time evolution of the bonding probability $p_b(t)$ of
a single bond, obtained from the particle-based simulations (symbols) and 
a least squares fit of Eq.~(\ref{pbt}) (solid lines). Clearly, the rate
equation Eq.~(\ref{pbt}) captures the behaviour of $p_b(t)$ accurately. In
addition, this procedure provides estimates of $\Gamma$ and $p_\infty$, which
through Eq.~(\ref{gamma}) yield estimates of the breaking and bonding rates,
$(k_{br},k_{bo})$, over a wide range of temperatures and densities. As shown
in Fig.~\ref{pbvsrho}, we find a remarkable agreement between
$p_\infty(\rho,T)$ obtained from the fit and $p_{b,eq}(\rho,T)$ obtained from
Wertheim's theory. The observed deviations at the lowest simulated
temperature may be due to the approximations required to calculate $\Delta$,
as the mapping of the reference system to a system of hard spheres with an
effective diameter, or the fact that, at high densities and low temperatures a 
gel is formed.

%%%%%%%%%%%%%%%%%%%%%%%%%%%%%%%
\begin{figure}
   \begin{center} 
   \includegraphics[width=\columnwidth]{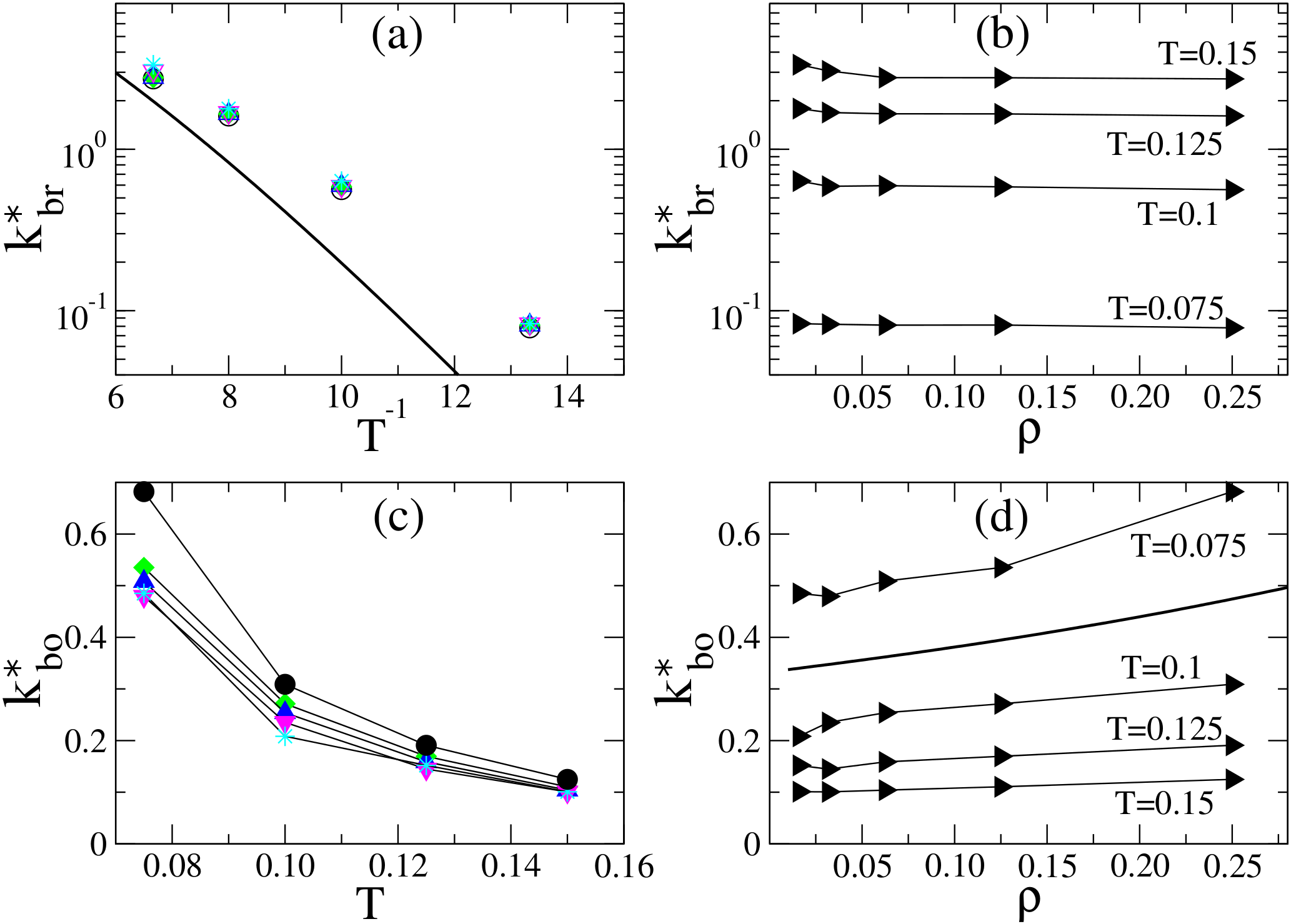} \\
   \caption{Temperature and density dependence of the asymptotic value of the
	   rate of bond breaking ($k^*_{br}$) and bonding ($k^*_{bo}$) of a
	   single bond. Rate of bond breaking of a single bond as a function
	   of (a) the inverse temperature and (b) density, at different
	   densities and temperatures, respectively. Rate of bonding of a
	   single bond as a function of (c) temperature and (d) density, at
	   different densities and temperatures, respectively. The solid lines
	   in (a) and (d), and the symbols in (a) and (c) have the same
	   meaning as in Fig.~\ref{kbreak}.~\label{kinf}}
\end{center}
\end{figure}
%%%%%%%%%%%%%%%%%%%%%%%%%%%%%%%

As discussed above,  it is
possible to estimate the temperature dependence of the effective rates
$k_{br}$ and $k_{bo}$, from $\Gamma$ (Fig.~\ref{pbvstime}) , as shown in
Fig.~\ref{kbreak}. We see that $k_{br}$ follows an exponential decay with
$1/T$, as expected for a thermal activated (Arrhenius) process, with no
significant dependence on the density. Following Arrhenius' work, we expect
that $k_{br}\sim\langle\exp(\beta V_G)\rangle$, where $\beta=(k_BT)^{-1}$ and
$\langle\cdots\rangle$ stands for an ensemble average over different bonds.
From Wertheim's theory, Eq.~(\ref{Delta2}), $G(T)=\langle\exp{(-\beta
V_G)}\rangle$, and we expect $k_{br}\sim G(T)^{-1}$.  The solid line in
Fig.~\ref{kbreak}(a) corresponds to $aG(T)^{-1}$, where $a$ is a constant
adjusted to fit the numerical data. One sees that the lower the temperature,
the better the agreement between the line and the numerical data. This result
indicates that $k_{br}=F(T)G(T)^{-1}$, where $F(T)$ is a decreasing function
of T. We note that most other studies of patchy particles consider a square well potential for
the patch-patch interaction. In those systems, the energy barrier is equal to
the energy of the bond ($E_b$) and thus $k_{br}(T)\sim\exp(\beta E_b)$, since
$E_b$ is the same for all bonds~\cite{Sciortino2009}.

At equilibrium $k_{bo}=k_{br}\Delta$ and thus $k_{bo}\approx
g_{HS}(r=d,\rho)F(T)$.  Under these conditions, since the dependence on T is
only through $F(T)$, the bonding rate of a single bond should decrease with
temperature, as shown in Fig.~\ref{kbreak}(c). In Fig.~\ref{kbreak}(d) we
plot the dependence of $k_{bo}$ on density at different temperatures. The
solid line is proportional to $g_{HS}$, showing that the density dependence is
well described by $g_{HS}$. Note that, the rate of bonding should depend on
the diffusion coefficient of the particles, which sets the time scale for two
particles in solution to collide, and thus increases with temperature. However,
as time is rescaled by the Brownian time, this effect is not apparent in the
results.

For simplicity, in line with previous works, we considered that the rates
$k_{br}$ and $k_{bo}$ are indeed constant, i.e. time independent.
Particle-based simulations allow us to measure these rates and evaluate this
assumption. To that end, we measured the number of new and broken bonds over
intervals of $5000$ iteration steps (between $0.075$ and $0.15$ in Brownian
time, depending on the temperature) and estimated $k_{br}(t)$ and $k_{bo}(t)$. 
Fig.~\ref{kvst} illustrates the time dependence of
these rates, at different temperatures and $\rho=0.25$. We find that both
rates increase with time and saturate at long times. However, while $k_{br}$
saturates rapidly at times of the order of the Brownian time, $k_{bo}$ takes
up to one order of magnitude longer to saturate.  The lower the temperature
the longer the time it takes to saturate. In the same figure, we also plotted the 
effective rates estimated from the fit discussed above (horizontal lines). It is
clear that the asymptotic values of these rates are significantly different from
the effective rates obtained from the fit. In Fig.~\ref{kinf} we plot the
asymptotic values of these rates, $k^*_{br}$ and $k^*_{bo}$ as a function of
temperature and density (computed from an average over the long-time 
regime of $k_{br}(t)$ and $k_{bo}(t)$ - see Fig.~\ref{kvst}). While, the 
qualitative dependence of these rates on $T$ and $\rho$ is consistent with 
that obtained from the fits (see Fig.~\ref{kbreak}), the values
are significantly different. Nevertheless, their ratio is the same, as expected 
 Wertheim's theory~(\ref{lma})~and~(\ref{Deltakbkb}), with $p_\infty=p_{b,eq}$.
Finally, Eq.~(\ref{pbt}) with $\Gamma$ obtained using the asymptotic values of
$k^*_{br}$ and $k^*_{bo}$ from the numerical simulations
does not fit the data of Fig.~\ref{pbvstime} (not shown). This
clearly suggests that the out of equilibrium dynamics is different from that
at equilibrium, at least in three dimensions, and that the effective rates
describing the initial dynamics are different from those that describe the
dynamics at equilibrium.

\section{Conclusions\label{sec.conclusions}}
We combined Langevin particle-based simulations, mean-field rate equations,
and Wertheim's equilibrium theory to study the dynamics of patchy particles
both in and out of equilibrium. By analyzing the dependence of the rate of
breaking and forming a single bond, at different temperatures and densities,
we have shown that the dynamics at equilibrium is different from that towards
equilibrium (out of equilibrium). In particular, for initial systems of
randomly distributed particles with no bonds, both rates are systematically
higher at equilibrium. However, since the asymptotic value of the
fraction of bonded patches depends on the ratio of the two rates, it
can be estimated both from the dynamics and the equilibrium theory.

We have focused on the non-percolative regime (high temperature and low
density). At low temperatures and high densities, percolation is expected to
occur and thus collective effects will definitely affect the
dynamics~\cite{Dias2016a}. Assumptions such as negligible patch-patch
correlations and independence of the diffusion coefficient of the cluster size
are no longer valid. In that regime, we expect differences between the
equilibrium and out of equilibrium dynamics to be more significant and the validity
of the mean-field approach compromised.

This work highlights the relevance of out of equilibrium studies. It was shown
previously that, under certain conditions, the dynamics depends strongly on
the initial configuration~\cite{Dias2016,Dias2016a}. Here, we have considered
only initially unbonded particles. The difference between the equilibrium and
out of equilibrium dynamics is expected to depend also on the initial conditions.
This difference is expected to impact most strongly on the rate of bonding, as
different kinetic pathways will correspond to different mechanisms of bond
formation. Notwithstanding, in the limit of very strong patch-patch
correlations, even the bond breaking dynamics may be affected by collective
effects.

\begin{acknowledgments}
We acknowledge financial support from the Portuguese Foundation for Science
and Technology (FCT) under Contracts nos. EXCL/FIS-NAN/0083/2012,
UID/FIS/00618/2013, IF/00255/2013, SFRH/BPD/114839/2016, and FCT/DAAD
bilateral project. 
\end{acknowledgments}

\bibliography{paper_JCCP}

\begin{thebibliography}{41}%
\makeatletter
\providecommand \@ifxundefined [1]{%
 \@ifx{#1\undefined}
}%
\providecommand \@ifnum [1]{%
 \ifnum #1\expandafter \@firstoftwo
 \else \expandafter \@secondoftwo
 \fi
}%
\providecommand \@ifx [1]{%
 \ifx #1\expandafter \@firstoftwo
 \else \expandafter \@secondoftwo
 \fi
}%
\providecommand \natexlab [1]{#1}%
\providecommand \enquote  [1]{``#1''}%
\providecommand \bibnamefont  [1]{#1}%
\providecommand \bibfnamefont [1]{#1}%
\providecommand \citenamefont [1]{#1}%
\providecommand \href@noop [0]{\@secondoftwo}%
\providecommand \href [0]{\begingroup \@sanitize@url \@href}%
\providecommand \@href[1]{\@@startlink{#1}\@@href}%
\providecommand \@@href[1]{\endgroup#1\@@endlink}%
\providecommand \@sanitize@url [0]{\catcode `\\12\catcode `\$12\catcode
  `\&12\catcode `\#12\catcode `\^12\catcode `\_12\catcode `\%12\relax}%
\providecommand \@@startlink[1]{}%
\providecommand \@@endlink[0]{}%
\providecommand \url  [0]{\begingroup\@sanitize@url \@url }%
\providecommand \@url [1]{\endgroup\@href {#1}{\urlprefix }}%
\providecommand \urlprefix  [0]{URL }%
\providecommand \Eprint [0]{\href }%
\providecommand \doibase [0]{http://dx.doi.org/}%
\providecommand \selectlanguage [0]{\@gobble}%
\providecommand \bibinfo  [0]{\@secondoftwo}%
\providecommand \bibfield  [0]{\@secondoftwo}%
\providecommand \translation [1]{[#1]}%
\providecommand \BibitemOpen [0]{}%
\providecommand \bibitemStop [0]{}%
\providecommand \bibitemNoStop [0]{.\EOS\space}%
\providecommand \EOS [0]{\spacefactor3000\relax}%
\providecommand \BibitemShut  [1]{\csname bibitem#1\endcsname}%
\let\auto@bib@innerbib\@empty
%</preamble>
\bibitem [{\citenamefont {Doppelbauer}\ \emph {et~al.}(2010)\citenamefont
  {Doppelbauer}, \citenamefont {Bianchi},\ and\ \citenamefont
  {Kahl}}]{Doppelbauer2010}%
  \BibitemOpen
  \bibfield  {author} {\bibinfo {author} {\bibfnamefont {G.}~\bibnamefont
  {Doppelbauer}}, \bibinfo {author} {\bibfnamefont {E.}~\bibnamefont
  {Bianchi}}, \ and\ \bibinfo {author} {\bibfnamefont {G.}~\bibnamefont
  {Kahl}},\ }\href@noop {} {\bibfield  {journal} {\bibinfo  {journal} {J.
  Phys.: Condens. Matter}\ }\textbf {\bibinfo {volume} {22}},\ \bibinfo {pages}
  {104105} (\bibinfo {year} {2010})}\BibitemShut {NoStop}%
\bibitem [{\citenamefont {Glotzer}\ and\ \citenamefont
  {Solomon}(2007)}]{Glotzer2007}%
  \BibitemOpen
  \bibfield  {author} {\bibinfo {author} {\bibfnamefont {S.~C.}\ \bibnamefont
  {Glotzer}}\ and\ \bibinfo {author} {\bibfnamefont {M.~J.}\ \bibnamefont
  {Solomon}},\ }\href@noop {} {\bibfield  {journal} {\bibinfo  {journal} {Nat.
  Mater.}\ }\textbf {\bibinfo {volume} {6}},\ \bibinfo {pages} {557} (\bibinfo
  {year} {2007})}\BibitemShut {NoStop}%
\bibitem [{\citenamefont {Frenkel}\ and\ \citenamefont
  {Wales}(2011)}]{Frenkel2011}%
  \BibitemOpen
  \bibfield  {author} {\bibinfo {author} {\bibfnamefont {D.}~\bibnamefont
  {Frenkel}}\ and\ \bibinfo {author} {\bibfnamefont {D.~J.}\ \bibnamefont
  {Wales}},\ }\href@noop {} {\bibfield  {journal} {\bibinfo  {journal} {Nat.
  Mater.}\ }\textbf {\bibinfo {volume} {10}},\ \bibinfo {pages} {410} (\bibinfo
  {year} {2011})}\BibitemShut {NoStop}%
\bibitem [{\citenamefont {Kufer}\ \emph {et~al.}(2008)\citenamefont {Kufer},
  \citenamefont {Puchner}, \citenamefont {Gumpp}, \citenamefont {Liedl},\ and\
  \citenamefont {Gaub}}]{Kufer2008}%
  \BibitemOpen
  \bibfield  {author} {\bibinfo {author} {\bibfnamefont {S.~K.}\ \bibnamefont
  {Kufer}}, \bibinfo {author} {\bibfnamefont {E.~M.}\ \bibnamefont {Puchner}},
  \bibinfo {author} {\bibfnamefont {H.}~\bibnamefont {Gumpp}}, \bibinfo
  {author} {\bibfnamefont {T.}~\bibnamefont {Liedl}}, \ and\ \bibinfo {author}
  {\bibfnamefont {H.~E.}\ \bibnamefont {Gaub}},\ }\href@noop {} {\bibfield
  {journal} {\bibinfo  {journal} {Science}\ }\textbf {\bibinfo {volume}
  {319}},\ \bibinfo {pages} {594} (\bibinfo {year} {2008})}\BibitemShut
  {NoStop}%
\bibitem [{\citenamefont {Whitesides}\ and\ \citenamefont
  {Grzybowski}(2002)}]{Whitesides2002}%
  \BibitemOpen
  \bibfield  {author} {\bibinfo {author} {\bibfnamefont {G.~M.}\ \bibnamefont
  {Whitesides}}\ and\ \bibinfo {author} {\bibfnamefont {B.}~\bibnamefont
  {Grzybowski}},\ }\href@noop {} {\bibfield  {journal} {\bibinfo  {journal}
  {Science}\ }\textbf {\bibinfo {volume} {295}},\ \bibinfo {pages} {2418}
  (\bibinfo {year} {2002})}\BibitemShut {NoStop}%
\bibitem [{\citenamefont {Ma}\ and\ \citenamefont {Hao}(2011)}]{Ma2011}%
  \BibitemOpen
  \bibfield  {author} {\bibinfo {author} {\bibfnamefont {H.}~\bibnamefont
  {Ma}}\ and\ \bibinfo {author} {\bibfnamefont {J.}~\bibnamefont {Hao}},\
  }\href@noop {} {\bibfield  {journal} {\bibinfo  {journal} {Chem. Soc. Rev.}\
  }\textbf {\bibinfo {volume} {40}},\ \bibinfo {pages} {5457} (\bibinfo {year}
  {2011})}\BibitemShut {NoStop}%
\bibitem [{\citenamefont {Zhang}\ \emph {et~al.}(2015)\citenamefont {Zhang},
  \citenamefont {Luijten},\ and\ \citenamefont {Granick}}]{Zhang2014a}%
  \BibitemOpen
  \bibfield  {author} {\bibinfo {author} {\bibfnamefont {J.}~\bibnamefont
  {Zhang}}, \bibinfo {author} {\bibfnamefont {E.}~\bibnamefont {Luijten}}, \
  and\ \bibinfo {author} {\bibfnamefont {S.}~\bibnamefont {Granick}},\
  }\href@noop {} {\bibfield  {journal} {\bibinfo  {journal} {Ann. Rev. Phys.
  Chem.}\ }\textbf {\bibinfo {volume} {66}},\ \bibinfo {pages} {581} (\bibinfo
  {year} {2015})}\BibitemShut {NoStop}%
\bibitem [{\citenamefont {Paulson}\ \emph {et~al.}(2015)\citenamefont
  {Paulson}, \citenamefont {Mesbah}, \citenamefont {Zhu}, \citenamefont
  {Molaro},\ and\ \citenamefont {Braatz}}]{Paulson2015}%
  \BibitemOpen
  \bibfield  {author} {\bibinfo {author} {\bibfnamefont {J.~A.}\ \bibnamefont
  {Paulson}}, \bibinfo {author} {\bibfnamefont {A.}~\bibnamefont {Mesbah}},
  \bibinfo {author} {\bibfnamefont {X.}~\bibnamefont {Zhu}}, \bibinfo {author}
  {\bibfnamefont {M.~C.}\ \bibnamefont {Molaro}}, \ and\ \bibinfo {author}
  {\bibfnamefont {R.~D.}\ \bibnamefont {Braatz}},\ }\href@noop {} {\bibfield
  {journal} {\bibinfo  {journal} {J. Proc. Cont.}\ }\textbf {\bibinfo {volume}
  {27}},\ \bibinfo {pages} {38} (\bibinfo {year} {2015})}\BibitemShut {NoStop}%
\bibitem [{\citenamefont {Lu}\ and\ \citenamefont {Weitz}(2013)}]{Lu2013}%
  \BibitemOpen
  \bibfield  {author} {\bibinfo {author} {\bibfnamefont {P.~J.}\ \bibnamefont
  {Lu}}\ and\ \bibinfo {author} {\bibfnamefont {D.~A.}\ \bibnamefont {Weitz}},\
  }\href@noop {} {\bibfield  {journal} {\bibinfo  {journal} {Annu. Rev.
  Condens. Matter Phys.}\ }\textbf {\bibinfo {volume} {4}},\ \bibinfo {pages}
  {217} (\bibinfo {year} {2013})}\BibitemShut {NoStop}%
\bibitem [{\citenamefont {Duguet}\ \emph {et~al.}(2011)\citenamefont {Duguet},
  \citenamefont {D{\'{e}}sert}, \citenamefont {Perro},\ and\ \citenamefont
  {Ravaine}}]{Duguet2011}%
  \BibitemOpen
  \bibfield  {author} {\bibinfo {author} {\bibfnamefont {E.}~\bibnamefont
  {Duguet}}, \bibinfo {author} {\bibfnamefont {A.}~\bibnamefont
  {D{\'{e}}sert}}, \bibinfo {author} {\bibfnamefont {A.}~\bibnamefont {Perro}},
  \ and\ \bibinfo {author} {\bibfnamefont {S.}~\bibnamefont {Ravaine}},\
  }\href@noop {} {\bibfield  {journal} {\bibinfo  {journal} {Chem. Soc. Rev.}\
  }\textbf {\bibinfo {volume} {40}},\ \bibinfo {pages} {941} (\bibinfo {year}
  {2011})}\BibitemShut {NoStop}%
\bibitem [{\citenamefont {Hu}\ \emph {et~al.}(2012)\citenamefont {Hu},
  \citenamefont {Zhou}, \citenamefont {Sun}, \citenamefont {Fang},\ and\
  \citenamefont {Wu}}]{Hu2012}%
  \BibitemOpen
  \bibfield  {author} {\bibinfo {author} {\bibfnamefont {J.}~\bibnamefont
  {Hu}}, \bibinfo {author} {\bibfnamefont {S.}~\bibnamefont {Zhou}}, \bibinfo
  {author} {\bibfnamefont {Y.}~\bibnamefont {Sun}}, \bibinfo {author}
  {\bibfnamefont {X.}~\bibnamefont {Fang}}, \ and\ \bibinfo {author}
  {\bibfnamefont {L.}~\bibnamefont {Wu}},\ }\href@noop {} {\bibfield  {journal}
  {\bibinfo  {journal} {Chem. Soc. Rev.}\ }\textbf {\bibinfo {volume} {41}},\
  \bibinfo {pages} {4356} (\bibinfo {year} {2012})}\BibitemShut {NoStop}%
\bibitem [{\citenamefont {Kretzschmar}\ and\ \citenamefont
  {Song}(2011)}]{Kretzschmar2011}%
  \BibitemOpen
  \bibfield  {author} {\bibinfo {author} {\bibfnamefont {I.}~\bibnamefont
  {Kretzschmar}}\ and\ \bibinfo {author} {\bibfnamefont {J.~H.}\ \bibnamefont
  {Song}},\ }\href@noop {} {\bibfield  {journal} {\bibinfo  {journal} {Curr.
  Op. Coll. Interf. Sci.}\ }\textbf {\bibinfo {volume} {16}},\ \bibinfo {pages}
  {84} (\bibinfo {year} {2011})}\BibitemShut {NoStop}%
\bibitem [{\citenamefont {Sacanna}\ and\ \citenamefont
  {Pine}(2011)}]{Sacanna2011}%
  \BibitemOpen
  \bibfield  {author} {\bibinfo {author} {\bibfnamefont {S.}~\bibnamefont
  {Sacanna}}\ and\ \bibinfo {author} {\bibfnamefont {D.~J.}\ \bibnamefont
  {Pine}},\ }\href@noop {} {\bibfield  {journal} {\bibinfo  {journal} {Curr.
  Op. Coll. Interf. Sci.}\ }\textbf {\bibinfo {volume} {16}},\ \bibinfo {pages}
  {96} (\bibinfo {year} {2011})}\BibitemShut {NoStop}%
\bibitem [{\citenamefont {Solomon}(2011)}]{Solomon2011}%
  \BibitemOpen
  \bibfield  {author} {\bibinfo {author} {\bibfnamefont {M.~J.}\ \bibnamefont
  {Solomon}},\ }\href@noop {} {\bibfield  {journal} {\bibinfo  {journal} {Curr.
  Op. Coll. Interf. Sci.}\ }\textbf {\bibinfo {volume} {16}},\ \bibinfo {pages}
  {158} (\bibinfo {year} {2011})}\BibitemShut {NoStop}%
\bibitem [{\citenamefont {Pawar}\ and\ \citenamefont
  {Kretzschmar}(2010)}]{Pawar2010}%
  \BibitemOpen
  \bibfield  {author} {\bibinfo {author} {\bibfnamefont {A.~B.}\ \bibnamefont
  {Pawar}}\ and\ \bibinfo {author} {\bibfnamefont {I.}~\bibnamefont
  {Kretzschmar}},\ }\href@noop {} {\bibfield  {journal} {\bibinfo  {journal}
  {Macromol. Rapid Commun.}\ }\textbf {\bibinfo {volume} {31}},\ \bibinfo
  {pages} {150} (\bibinfo {year} {2010})}\BibitemShut {NoStop}%
\bibitem [{\citenamefont {Sacanna}\ \emph {et~al.}(2013)\citenamefont
  {Sacanna}, \citenamefont {Pine},\ and\ \citenamefont {Yi}}]{Sacanna2013a}%
  \BibitemOpen
  \bibfield  {author} {\bibinfo {author} {\bibfnamefont {S.}~\bibnamefont
  {Sacanna}}, \bibinfo {author} {\bibfnamefont {D.~J.}\ \bibnamefont {Pine}}, \
  and\ \bibinfo {author} {\bibfnamefont {G.-R.}\ \bibnamefont {Yi}},\
  }\href@noop {} {\bibfield  {journal} {\bibinfo  {journal} {Soft Matt.}\
  }\textbf {\bibinfo {volume} {9}},\ \bibinfo {pages} {8096} (\bibinfo {year}
  {2013})}\BibitemShut {NoStop}%
\bibitem [{\citenamefont {Manoharan}(2015)}]{Manoharan2015}%
  \BibitemOpen
  \bibfield  {author} {\bibinfo {author} {\bibfnamefont {V.~N.}\ \bibnamefont
  {Manoharan}},\ }\href@noop {} {\bibfield  {journal} {\bibinfo  {journal}
  {Science}\ }\textbf {\bibinfo {volume} {349}},\ \bibinfo {pages} {1253751}
  (\bibinfo {year} {2015})}\BibitemShut {NoStop}%
\bibitem [{\citenamefont {Zaccarelli}(2007)}]{Zaccarelli2007}%
  \BibitemOpen
  \bibfield  {author} {\bibinfo {author} {\bibfnamefont {E.}~\bibnamefont
  {Zaccarelli}},\ }\href@noop {} {\bibfield  {journal} {\bibinfo  {journal} {J.
  Phys.: Condens. Matter}\ }\textbf {\bibinfo {volume} {19}},\ \bibinfo {pages}
  {323101} (\bibinfo {year} {2007})}\BibitemShut {NoStop}%
\bibitem [{\citenamefont {Dias}\ \emph {et~al.}(2017)\citenamefont {Dias},
  \citenamefont {Ara{\'{u}}jo},\ and\ \citenamefont {{Telo da
  Gama}}}]{Dias2017}%
  \BibitemOpen
  \bibfield  {author} {\bibinfo {author} {\bibfnamefont {C.~S.}\ \bibnamefont
  {Dias}}, \bibinfo {author} {\bibfnamefont {N.~A.~M.}\ \bibnamefont
  {Ara{\'{u}}jo}}, \ and\ \bibinfo {author} {\bibfnamefont {M.~M.}\
  \bibnamefont {{Telo da Gama}}},\ }\href@noop {} {\bibfield  {journal}
  {\bibinfo  {journal} {Adv. Col. Interf. Sci.}\ }\textbf {\bibinfo {volume}
  {247}},\ \bibinfo {pages} {258} (\bibinfo {year} {2017})}\BibitemShut
  {NoStop}%
\bibitem [{\citenamefont {Zaccarelli}\ \emph {et~al.}(2006)\citenamefont
  {Zaccarelli}, \citenamefont {Saika-Voivod}, \citenamefont {Buldyrev},
  \citenamefont {Moreno}, \citenamefont {Tartaglia},\ and\ \citenamefont
  {Sciortino}}]{Zaccarelli2006}%
  \BibitemOpen
  \bibfield  {author} {\bibinfo {author} {\bibfnamefont {E.}~\bibnamefont
  {Zaccarelli}}, \bibinfo {author} {\bibfnamefont {I.}~\bibnamefont
  {Saika-Voivod}}, \bibinfo {author} {\bibfnamefont {S.~V.}\ \bibnamefont
  {Buldyrev}}, \bibinfo {author} {\bibfnamefont {A.~J.}\ \bibnamefont
  {Moreno}}, \bibinfo {author} {\bibfnamefont {P.}~\bibnamefont {Tartaglia}}, \
  and\ \bibinfo {author} {\bibfnamefont {F.}~\bibnamefont {Sciortino}},\
  }\href@noop {} {\bibfield  {journal} {\bibinfo  {journal} {J. Chem. Phys.}\
  }\textbf {\bibinfo {volume} {124}},\ \bibinfo {pages} {124908} (\bibinfo
  {year} {2006})}\BibitemShut {NoStop}%
\bibitem [{\citenamefont {Dias}\ \emph
  {et~al.}(2013{\natexlab{a}})\citenamefont {Dias}, \citenamefont
  {Ara{\'{u}}jo},\ and\ \citenamefont {{Telo da Gama}}}]{Dias2013}%
  \BibitemOpen
  \bibfield  {author} {\bibinfo {author} {\bibfnamefont {C.~S.}\ \bibnamefont
  {Dias}}, \bibinfo {author} {\bibfnamefont {N.~A.~M.}\ \bibnamefont
  {Ara{\'{u}}jo}}, \ and\ \bibinfo {author} {\bibfnamefont {M.~M.}\
  \bibnamefont {{Telo da Gama}}},\ }\href@noop {} {\bibfield  {journal}
  {\bibinfo  {journal} {Phys. Rev. E}\ }\textbf {\bibinfo {volume} {87}},\
  \bibinfo {pages} {032308} (\bibinfo {year} {2013}{\natexlab{a}})}\BibitemShut
  {NoStop}%
\bibitem [{\citenamefont {Dias}\ \emph
  {et~al.}(2013{\natexlab{b}})\citenamefont {Dias}, \citenamefont
  {Ara{\'{u}}jo},\ and\ \citenamefont {{Telo da Gama}}}]{Dias2013a}%
  \BibitemOpen
  \bibfield  {author} {\bibinfo {author} {\bibfnamefont {C.~S.}\ \bibnamefont
  {Dias}}, \bibinfo {author} {\bibfnamefont {N.~A.~M.}\ \bibnamefont
  {Ara{\'{u}}jo}}, \ and\ \bibinfo {author} {\bibfnamefont {M.~M.}\
  \bibnamefont {{Telo da Gama}}},\ }\href@noop {} {\bibfield  {journal}
  {\bibinfo  {journal} {Soft Matt.}\ }\textbf {\bibinfo {volume} {9}},\
  \bibinfo {pages} {5616} (\bibinfo {year} {2013}{\natexlab{b}})}\BibitemShut
  {NoStop}%
\bibitem [{\citenamefont {Chakrabarti}\ \emph {et~al.}(2014)\citenamefont
  {Chakrabarti}, \citenamefont {Kusumaatmaja}, \citenamefont {R{\"{u}}hle},\
  and\ \citenamefont {Wales}}]{Chakrabarti2014}%
  \BibitemOpen
  \bibfield  {author} {\bibinfo {author} {\bibfnamefont {D.}~\bibnamefont
  {Chakrabarti}}, \bibinfo {author} {\bibfnamefont {H.}~\bibnamefont
  {Kusumaatmaja}}, \bibinfo {author} {\bibfnamefont {V.}~\bibnamefont
  {R{\"{u}}hle}}, \ and\ \bibinfo {author} {\bibfnamefont {D.~J.}\ \bibnamefont
  {Wales}},\ }\href@noop {} {\bibfield  {journal} {\bibinfo  {journal} {Phys.
  Chem. Chem. Phys.}\ }\textbf {\bibinfo {volume} {16}},\ \bibinfo {pages}
  {5014} (\bibinfo {year} {2014})}\BibitemShut {NoStop}%
\bibitem [{\citenamefont {Dias}\ \emph {et~al.}(2014)\citenamefont {Dias},
  \citenamefont {Ara{\'{u}}jo},\ and\ \citenamefont {{Telo da
  Gama}}}]{Dias2014}%
  \BibitemOpen
  \bibfield  {author} {\bibinfo {author} {\bibfnamefont {C.~S.}\ \bibnamefont
  {Dias}}, \bibinfo {author} {\bibfnamefont {N.~A.~M.}\ \bibnamefont
  {Ara{\'{u}}jo}}, \ and\ \bibinfo {author} {\bibfnamefont {M.~M.}\
  \bibnamefont {{Telo da Gama}}},\ }\href@noop {} {\bibfield  {journal}
  {\bibinfo  {journal} {Phys. Rev. E}\ }\textbf {\bibinfo {volume} {90}},\
  \bibinfo {pages} {032302} (\bibinfo {year} {2014})}\BibitemShut {NoStop}%
\bibitem [{\citenamefont {Dias}\ \emph {et~al.}(2015)\citenamefont {Dias},
  \citenamefont {Ara{\'{u}}jo},\ and\ \citenamefont {{Telo da
  Gama}}}]{Dias2015}%
  \BibitemOpen
  \bibfield  {author} {\bibinfo {author} {\bibfnamefont {C.~S.}\ \bibnamefont
  {Dias}}, \bibinfo {author} {\bibfnamefont {N.~A.~M.}\ \bibnamefont
  {Ara{\'{u}}jo}}, \ and\ \bibinfo {author} {\bibfnamefont {M.~M.}\
  \bibnamefont {{Telo da Gama}}},\ }\href@noop {} {\bibfield  {journal}
  {\bibinfo  {journal} {Mol. Phys.}\ }\textbf {\bibinfo {volume} {113}},\
  \bibinfo {pages} {1069} (\bibinfo {year} {2015})}\BibitemShut {NoStop}%
\bibitem [{\citenamefont {Ara{\'{u}}jo}\ \emph {et~al.}(2015)\citenamefont
  {Ara{\'{u}}jo}, \citenamefont {Dias},\ and\ \citenamefont {{Telo da
  Gama}}}]{Araujo2015}%
  \BibitemOpen
  \bibfield  {author} {\bibinfo {author} {\bibfnamefont {N.~A.~M.}\
  \bibnamefont {Ara{\'{u}}jo}}, \bibinfo {author} {\bibfnamefont {C.~S.}\
  \bibnamefont {Dias}}, \ and\ \bibinfo {author} {\bibfnamefont {M.~M.}\
  \bibnamefont {{Telo da Gama}}},\ }\href@noop {} {\bibfield  {journal}
  {\bibinfo  {journal} {J. Phys.: Condens. Matter}\ }\textbf {\bibinfo {volume}
  {27}},\ \bibinfo {pages} {194123} (\bibinfo {year} {2015})}\BibitemShut
  {NoStop}%
\bibitem [{\citenamefont {Kartha}\ and\ \citenamefont
  {Sayeed}(2016)}]{Kartha2016}%
  \BibitemOpen
  \bibfield  {author} {\bibinfo {author} {\bibfnamefont {M.~J.}\ \bibnamefont
  {Kartha}}\ and\ \bibinfo {author} {\bibfnamefont {A.}~\bibnamefont
  {Sayeed}},\ }\href@noop {} {\bibfield  {journal} {\bibinfo  {journal} {Phys.
  Lett. A}\ }\textbf {\bibinfo {volume} {380}},\ \bibinfo {pages} {2791}
  (\bibinfo {year} {2016})}\BibitemShut {NoStop}%
\bibitem [{\citenamefont {Dias}\ \emph {et~al.}(2016)\citenamefont {Dias},
  \citenamefont {Braga}, \citenamefont {Ara{\'{u}}jo},\ and\ \citenamefont
  {{Telo da Gama}}}]{Dias2016}%
  \BibitemOpen
  \bibfield  {author} {\bibinfo {author} {\bibfnamefont {C.~S.}\ \bibnamefont
  {Dias}}, \bibinfo {author} {\bibfnamefont {C.}~\bibnamefont {Braga}},
  \bibinfo {author} {\bibfnamefont {N.~A.~M.}\ \bibnamefont {Ara{\'{u}}jo}}, \
  and\ \bibinfo {author} {\bibfnamefont {M.~M.}\ \bibnamefont {{Telo da
  Gama}}},\ }\href@noop {} {\bibfield  {journal} {\bibinfo  {journal} {Soft
  Matt.}\ }\textbf {\bibinfo {volume} {12}},\ \bibinfo {pages} {1550} (\bibinfo
  {year} {2016})}\BibitemShut {NoStop}%
\bibitem [{\citenamefont {Kartha}(2016)}]{Kartha2016a}%
  \BibitemOpen
  \bibfield  {author} {\bibinfo {author} {\bibfnamefont {M.~J.}\ \bibnamefont
  {Kartha}},\ }\href@noop {} {\bibfield  {journal} {\bibinfo  {journal} {Phys.
  Lett. A}\ }\textbf {\bibinfo {volume} {381}},\ \bibinfo {pages} {556}
  (\bibinfo {year} {2016})}\BibitemShut {NoStop}%
\bibitem [{\citenamefont {Ara{\'{u}}jo}\ \emph {et~al.}(2017)\citenamefont
  {Ara{\'{u}}jo}, \citenamefont {Dias},\ and\ \citenamefont {{Telo da
  Gama}}}]{Araujo2017}%
  \BibitemOpen
  \bibfield  {author} {\bibinfo {author} {\bibfnamefont {N.~A.~M.}\
  \bibnamefont {Ara{\'{u}}jo}}, \bibinfo {author} {\bibfnamefont {C.~S.}\
  \bibnamefont {Dias}}, \ and\ \bibinfo {author} {\bibfnamefont {M.~M.}\
  \bibnamefont {{Telo da Gama}}},\ }\href@noop {} {\bibfield  {journal}
  {\bibinfo  {journal} {J. Phys.: Condens. Matter}\ }\textbf {\bibinfo {volume}
  {29}},\ \bibinfo {pages} {014001} (\bibinfo {year} {2017})}\BibitemShut
  {NoStop}%
\bibitem [{\citenamefont {Corezzi}\ \emph {et~al.}(2010)\citenamefont
  {Corezzi}, \citenamefont {Fioretto}, \citenamefont {{De Michele}},
  \citenamefont {Zaccarelli},\ and\ \citenamefont {Sciortino}}]{Corezzi2010}%
  \BibitemOpen
  \bibfield  {author} {\bibinfo {author} {\bibfnamefont {S.}~\bibnamefont
  {Corezzi}}, \bibinfo {author} {\bibfnamefont {D.}~\bibnamefont {Fioretto}},
  \bibinfo {author} {\bibfnamefont {C.}~\bibnamefont {{De Michele}}}, \bibinfo
  {author} {\bibfnamefont {E.}~\bibnamefont {Zaccarelli}}, \ and\ \bibinfo
  {author} {\bibfnamefont {F.}~\bibnamefont {Sciortino}},\ }\href@noop {}
  {\bibfield  {journal} {\bibinfo  {journal} {J. Phys. Chem. B}\ }\textbf
  {\bibinfo {volume} {114}},\ \bibinfo {pages} {3769} (\bibinfo {year}
  {2010})}\BibitemShut {NoStop}%
\bibitem [{\citenamefont {Corezzi}\ \emph {et~al.}(2009)\citenamefont
  {Corezzi}, \citenamefont {{De Michele}}, \citenamefont {Zaccarelli},
  \citenamefont {Tartaglia},\ and\ \citenamefont {Sciortino}}]{Corezzi2009}%
  \BibitemOpen
  \bibfield  {author} {\bibinfo {author} {\bibfnamefont {S.}~\bibnamefont
  {Corezzi}}, \bibinfo {author} {\bibfnamefont {C.}~\bibnamefont {{De
  Michele}}}, \bibinfo {author} {\bibfnamefont {E.}~\bibnamefont {Zaccarelli}},
  \bibinfo {author} {\bibfnamefont {P.}~\bibnamefont {Tartaglia}}, \ and\
  \bibinfo {author} {\bibfnamefont {F.}~\bibnamefont {Sciortino}},\ }\href@noop
  {} {\bibfield  {journal} {\bibinfo  {journal} {J. Phys. Chem. B}\ }\textbf
  {\bibinfo {volume} {113}},\ \bibinfo {pages} {1233} (\bibinfo {year}
  {2009})}\BibitemShut {NoStop}%
\bibitem [{\citenamefont {Sciortino}\ \emph {et~al.}(2009)\citenamefont
  {Sciortino}, \citenamefont {{De Michele}}, \citenamefont {Corezzi},
  \citenamefont {Russo}, \citenamefont {Zaccarelli},\ and\ \citenamefont
  {Tartaglia}}]{Sciortino2009}%
  \BibitemOpen
  \bibfield  {author} {\bibinfo {author} {\bibfnamefont {F.}~\bibnamefont
  {Sciortino}}, \bibinfo {author} {\bibfnamefont {C.}~\bibnamefont {{De
  Michele}}}, \bibinfo {author} {\bibfnamefont {S.}~\bibnamefont {Corezzi}},
  \bibinfo {author} {\bibfnamefont {J.}~\bibnamefont {Russo}}, \bibinfo
  {author} {\bibfnamefont {E.}~\bibnamefont {Zaccarelli}}, \ and\ \bibinfo
  {author} {\bibfnamefont {P.}~\bibnamefont {Tartaglia}},\ }\href@noop {}
  {\bibfield  {journal} {\bibinfo  {journal} {Soft Matt.}\ }\textbf {\bibinfo
  {volume} {5}},\ \bibinfo {pages} {2571} (\bibinfo {year} {2009})}\BibitemShut
  {NoStop}%
\bibitem [{\citenamefont {Dias}\ \emph {et~al.}({\natexlab{a}})\citenamefont
  {Dias}, \citenamefont {Tavares}, \citenamefont {Ara{\'{u}}jo},\ and\
  \citenamefont {{Telo da Gama}}}]{Dias2016a}%
  \BibitemOpen
  \bibfield  {author} {\bibinfo {author} {\bibfnamefont {C.~S.}\ \bibnamefont
  {Dias}}, \bibinfo {author} {\bibfnamefont {J.~M.}\ \bibnamefont {Tavares}},
  \bibinfo {author} {\bibfnamefont {N.~A.~M.}\ \bibnamefont {Ara{\'{u}}jo}}, \
  and\ \bibinfo {author} {\bibfnamefont {M.~M.}\ \bibnamefont {{Telo da
  Gama}}},\ }\href@noop {} {\ ,\ \bibinfo {pages} {arxiv: 1604.05279}
  ({\natexlab{a}})}\BibitemShut {NoStop}%
\bibitem [{\citenamefont {Dias}\ \emph {et~al.}({\natexlab{b}})\citenamefont
  {Dias}, \citenamefont {Ara{\'{u}}jo},\ and\ \citenamefont {{Telo da
  Gama}}}]{Dias2017b}%
  \BibitemOpen
  \bibfield  {author} {\bibinfo {author} {\bibfnamefont {C.~S.}\ \bibnamefont
  {Dias}}, \bibinfo {author} {\bibfnamefont {N.~A.~M.}\ \bibnamefont
  {Ara{\'{u}}jo}}, \ and\ \bibinfo {author} {\bibfnamefont {M.~M.}\
  \bibnamefont {{Telo da Gama}}},\ }\href@noop {} {\ ,\ \bibinfo {pages}
  {arXiv:1710.02373} ({\natexlab{b}})}\BibitemShut {NoStop}%
\bibitem [{\citenamefont {Vasilyev}\ \emph {et~al.}(2013)\citenamefont
  {Vasilyev}, \citenamefont {Klumov},\ and\ \citenamefont
  {Tkachenko}}]{Vasilyev2013}%
  \BibitemOpen
  \bibfield  {author} {\bibinfo {author} {\bibfnamefont {O.~A.}\ \bibnamefont
  {Vasilyev}}, \bibinfo {author} {\bibfnamefont {B.~A.}\ \bibnamefont
  {Klumov}}, \ and\ \bibinfo {author} {\bibfnamefont {A.~V.}\ \bibnamefont
  {Tkachenko}},\ }\href@noop {} {\bibfield  {journal} {\bibinfo  {journal}
  {Phys. Rev. E}\ }\textbf {\bibinfo {volume} {88}},\ \bibinfo {pages} {012302}
  (\bibinfo {year} {2013})}\BibitemShut {NoStop}%
\bibitem [{\citenamefont {Plimpton}(1995)}]{Plimpton1995}%
  \BibitemOpen
  \bibfield  {author} {\bibinfo {author} {\bibfnamefont {S.}~\bibnamefont
  {Plimpton}},\ }\href@noop {} {\bibfield  {journal} {\bibinfo  {journal} {J.
  Comp. Phys.}\ }\textbf {\bibinfo {volume} {117}},\ \bibinfo {pages} {1}
  (\bibinfo {year} {1995})}\BibitemShut {NoStop}%
\bibitem [{\citenamefont {Mazza}\ \emph {et~al.}(2007)\citenamefont {Mazza},
  \citenamefont {Giovambattista}, \citenamefont {Stanley},\ and\ \citenamefont
  {Starr}}]{Mazza2007}%
  \BibitemOpen
  \bibfield  {author} {\bibinfo {author} {\bibfnamefont {M.~G.}\ \bibnamefont
  {Mazza}}, \bibinfo {author} {\bibfnamefont {N.}~\bibnamefont
  {Giovambattista}}, \bibinfo {author} {\bibfnamefont {H.~E.}\ \bibnamefont
  {Stanley}}, \ and\ \bibinfo {author} {\bibfnamefont {F.~W.}\ \bibnamefont
  {Starr}},\ }\href@noop {} {\bibfield  {journal} {\bibinfo  {journal} {Phys.
  Rev. E}\ }\textbf {\bibinfo {volume} {76}},\ \bibinfo {pages} {031203}
  (\bibinfo {year} {2007})}\BibitemShut {NoStop}%
\bibitem [{\citenamefont {van Dongen}\ and\ \citenamefont
  {Ernst}(1984)}]{VanDongen1984}%
  \BibitemOpen
  \bibfield  {author} {\bibinfo {author} {\bibfnamefont {P.~G.~J.}\
  \bibnamefont {van Dongen}}\ and\ \bibinfo {author} {\bibfnamefont {M.~H.}\
  \bibnamefont {Ernst}},\ }\href@noop {} {\bibfield  {journal} {\bibinfo
  {journal} {J. Stat. Phys.}\ }\textbf {\bibinfo {volume} {37}},\ \bibinfo
  {pages} {301} (\bibinfo {year} {1984})}\BibitemShut {NoStop}%
\bibitem [{\citenamefont {Teixeira}\ and\ \citenamefont
  {Tavares}(2017)}]{Teixeira2017a}%
  \BibitemOpen
  \bibfield  {author} {\bibinfo {author} {\bibfnamefont {P.~I.~C.}\
  \bibnamefont {Teixeira}}\ and\ \bibinfo {author} {\bibfnamefont {J.~M.}\
  \bibnamefont {Tavares}},\ }\href@noop {} {\bibfield  {journal} {\bibinfo
  {journal} {Curr. Op. Coll. Interf. Sci.}\ }\textbf {\bibinfo {volume} {30}},\
  \bibinfo {pages} {16} (\bibinfo {year} {2017})}\BibitemShut {NoStop}%
\bibitem [{\citenamefont {Bianchi}\ \emph {et~al.}(2008)\citenamefont
  {Bianchi}, \citenamefont {Tartaglia}, \citenamefont {Zaccarelli},\ and\
  \citenamefont {Sciortino}}]{Bianchi2008}%
  \BibitemOpen
  \bibfield  {author} {\bibinfo {author} {\bibfnamefont {E.}~\bibnamefont
  {Bianchi}}, \bibinfo {author} {\bibfnamefont {P.}~\bibnamefont {Tartaglia}},
  \bibinfo {author} {\bibfnamefont {E.}~\bibnamefont {Zaccarelli}}, \ and\
  \bibinfo {author} {\bibfnamefont {F.}~\bibnamefont {Sciortino}},\ }\href@noop
  {} {\bibfield  {journal} {\bibinfo  {journal} {J. Chem. Phys.}\ }\textbf
  {\bibinfo {volume} {128}},\ \bibinfo {pages} {144504} (\bibinfo {year}
  {2008})}\BibitemShut {NoStop}%
\end{thebibliography}%

\end{document}